\documentclass[useAMS,usenatbib]{mn2e}
\usepackage{graphicx}
\usepackage{txfonts}
\title[A broad-band analysis of eight radio loud type 1 AGN selected in the hard X-ray band]
  {A broad-band spectral analysis of eight radio loud type 1 AGN selected in the hard X-ray band}
\author[M. Molina et al.]
{M.~Molina,$^1$ L.~Bassani,$^2$ A.~Malizia,$^2$ A.J.~Bird,$^1$ A.J.~Dean,$^1$ M.~Fiocchi,$^3$ F.~Panessa,$^3$  \\
\newauthor
A.~De Rosa,$^3$ R.~Landi$^2$\\
$^1$School of Physics and Astronomy, University of Southampton,
        SO17 1BJ, Southampton, U.K., \\
$^2$IASF/INAF, via Gobetti 101, I-40129 Bologna, Italy,\\
$^3$IASF/INAF, via Fosso del Cavaliere 100, I-00133 Rome, Italy}

\begin{document}

\date{}

\pagerange{\pageref{firstpage}--\pageref{lastpage}} \pubyear{2008}

\maketitle

\label{firstpage}
        
\begin{abstract}

Starting from a complete sample of type I AGN observed by \emph{INTEGRAL} in the 20-40 keV band, we have
selected a set of 8 AGN which can be classified as radio loud objects according to their 1.4 GHz power density,  radio to hard X-ray flux flux density ratio and  radio morphology. The sample
contains 6 Broad Line Radio Galaxies and 2 candidate ones. Most of the objects in our sample display a double lobe morphology,
both on small and large scales. For all the objects, we present broad-band (1-110 keV) spectral analysis using \emph{INTEGRAL} observations together with archival
\emph{XMM-Newton}, \emph{Chandra}, \emph{Swift/XRT} and \emph{Swift/BAT} data. We constrain the primary continuum (photon index and cut-off energy), intrinsic absorption and
reprocessing features (iron line and reflection) in  most of the objects. The sources analysed here show remarkable similarities to radio quiet type I AGN with respect to most of the parameters analysed; we only find marginal evidence for weaker reprocessing features in our objects compared to their radio quiet counterparts. Similarly we do not
find any correlation between the spectral parameters studied and the source core dominance or radio to 20-100 keV flux density ratios, suggesting that what makes our objects  radio loud has no effect on their high energy characteristics.

\end{abstract}


\section{Introduction}

Most of what is known about Active Galactic Nuclei (AGN) is essentially based on studies of radio quiet sources, which make up almost 90\% of the entire AGN population.
It is now widely accepted that active galaxies are powered by accretion onto a supermassive black hole: the observed radiation, spanning the entire electromagnetic spectrum, is produced by a cold accretion disk and by a hot corona, as proposed in the so-called two-phase model \citep{b15}. In the X-ray domain, the emission of radio quiet Broad
Line AGN (Seyfert 1) is, to the first order, well described by a power law of photon index 1.8-2.0, extending from a few keV to over 100 keV; at higher energies there is
evidence for an exponential cut-off, the exact value of which is still uncertain \citep{b22,b23}. Secondary features such as the Fe K$\alpha$ line and the Compton reflection
component are also commonly observed; they are considered to be the effects of reprocessing of the primary continuum and are relatively well understood \citep{b19}. 
Studies
performed on samples of Broad Line Radio Galaxies (BLRG from now on; \citealt{b24,b7}) have shown that these objects show optical/UV continuum and emission line
characteristics similar to their radio quiet counterparts, but  display some fundamental differences in their X-ray behaviour. BLRG, in fact, exhibit flatter/harder power
law slopes than radio quiet Seyfert 1 galaxies and are also known to have weaker reprocessing features  (e.g. \citealt{b24} and \citealt{b7}). The origin of these
differences is however far from being understood. A possible cause for the observed diversity could be ascribed to a different disk geometry and/or accretion flow
efficiencies, or to the presence of jets and beaming effects that contaminate/dilute the AGN component and the reprocessing features; ionised reflection, which naturally
produces weaker reflection features, has also been considered in the literature \citep{b60}. 
Now that a large sample of AGN detected above 20 keV by \emph{INTEGRAL} is available, it is possible and
important to perform a comparison between different classes of AGN, by measuring the shape of the primary continuum together with the high energy cut-off
and the reflection component. In the present work we focus on the broad-band (1-110 keV) spectral analysis of a sample of eight radio loud type 1 AGN, combining
\emph{XMM-Newton}, \emph{Chandra}, \emph{Swift/XRT} and \emph{BAT} data together with \emph{INTEGRAL/ISGRI} measurements.
We also compare our results to a sample of radio quiet Seyfert 1s detected by \emph{INTEGRAL} and recently studied by \citet{b21} and finally discuss the implications of our findings.

\section{The Sample}

The definition of radio loudness is rather vague, with different criteria applied in the literature (see \citealt{b50} for a critical discussion). Traditionally, empirical boundaries were set by the radio power density at 5 GHz (P$_{5GHz}$; \citealt{b51})  and the radio (at the same frequency) to optical B band flux density ratio (R${_B}$; \citealt{b52}): according to these boundaries, an AGN is radio loud if P$_{5GHz}$$\geq$10$^{32}$ erg s$^{-1}$ Hz$^{-1}$ and Log(R$_B$)$\geq$1. In time, different surrogate definitions of both parameters have appeared in the literature involving radio data at different frequencies and comparison with even UV and X-ray fluxes to deal with objects selected in bands other than the optical. More recently \citet{b50} suggested the use of the source radio morphology as a further criterion to divide radio loud from radio quiet objects. Since FR I morphology is quite rare in broad line radio loud AGN, while double-lobe FR II appearance is much more common, these authors suggest the adoption of this last morphology as a further discriminator, i.e. FR II-like objects are by default radio loud. 
We therefore selected our radio loud AGN on the basis of similar criteria: the power density at 1.4 GHz (P$_{1.4GHz}$), the radio (at the same frequency) to the 20-100 keV flux density ratio (R$_{HX}$) and the source morphology, again at 1.4 GHz. Sources were selected
from a set of 35 Broad Line (BL) galaxies extracted from a complete sample of \emph{INTEGRAL} detected AGN\footnote{This sample extracted from the 3$^{rd}$ IBIS/ISGRI catalogue \citep{b3,b49} contains all AGN detected with a significance greater than 5.5$\sigma$ in the 20-40 keV band and having z$\le$0.14; it contains 35 type 1 objects, 30 type 2 sources, 3 Narrow Line Seyfert 1s, and 7 Blazars/QSOs}. Because many of these objects are newly discovered, it is difficult to have information for all of them  at 5 GHz and/or in the B band, hence the use of the more ready available 1.4 GHz and  20-100 keV data. 1.4 GHz information has been obtained from the HEASARC database\footnote{http://heasarc.gsfc.nasa.gov/} for 29 objects in the sample, except in the case of IGR J13109-5552, where the 1.4 GHz flux was extrapolated using the 4.85 and 0.83 GHz fluxes available in the HEASARC radio catalogues.  \emph{INTEGRAL} data were derived from the fluxes reported in Bird et al. (2007; see also table 1A) and are available for all objects in the sample.  Candidate radio loud objects were selected on the basis of their location in the diagram shown in figure \ref{radio_bright}, where the radio quiet/loud boundaries on both axes are set by the weakest FR II source in our sample, i.e. IGR J21247+5058. This source has been fully discussed by \citet{b18} and it is now a well established case of a radio loud AGN on the basis of various arguments. In particular it has been classified as an FR II galaxy by \citet{b18} with a L$_{5GHz}$$\sim$10$^{31}$ erg s$^{-1}$Hz$^{-1}$, an order of magnitude lower than the traditional limit, making this source one of the weakest FR II galaxies reported to date.

By means of these boundaries we identified 8 radio loud AGN: QSO B0241+62, B3 0309+411, 3C 111, 3C 390.3, IGR J13109-5552, 4C 74.26, S5 2116+81 and IGR J21247+5058 (see table 1); note the ambiguous location of two additional objects, Markarian 6 and MCG+08-11-11, which are below the adopted boundaries but very close to IGR J21247+5058. Information at 5 GHz and in the B band are available for all these sources so that a cross-check can be made on their radio power density and radio-to-optical flux density ratio according to the more conventional definition used by \citet{b51} and \citet{b52}. Both values are listed in table 1 for all 8 candidates plus Markarian 6 and MCG+08-11-011, which we keep under consideration due to their location in figure \ref{radio_bright}.  5 AGN have P$_{5GHz}$$\geq$10$^{32}$ erg s$^{-1}$ Hz$^{-1}$ and in any case all but Markarian 6 and MCG+08-11-011 have a 5 GHz power density exceeding that of IGR J21247+5058. All the objects, including Mrk 6 and MCG+08-11-011, have Log (R$_B$)$\geq$1; note that in IGR J21247+5058, R$_B$ cannot be estimated due to the lack of information on the B magnitude.
 
Next we examined the radio morphology of all our 8 radio loud candidate AGN, plus again Mrk 6 and MCG+08-11-011 (see table 1). All but 2 sources (i.e. QSO B0241+62 and IGR J13109-5552) are well known Broad Line Radio Galaxies (see table 1 and NED, \citealt{b2} and \citealt{b18}) and show a morphology with two lobes extending from the central nucleus (see for example NVSS contour maps available in NED); the extension of the lobes for B3 0309+411 and 4C 74.26 is more than 1 Mpc, hence their classification as giant radio galaxies \citep{b29,b30}. These double-lobed objects are classified in the literature as FR II galaxies, with the exception of 4C 74.26 and S5 2116+81, which still have an uncertain/unknown nature. QSO B0241+62 and IGR J13109-5552 have been poorly studied at radio frequencies and their classification is uncertain; both are Seyfert 1 according to NED and \citealt{b59}. QSO B0241+62, shows a compact unresolved core in the NVSS map, although high resolution radio imaging indicates a possible double-lobe morphology within this core structure \citep{b27}. The sky region containing IGR J13109-5552 has not been mapped by the NVSS, but the source is present in the Molonglo Galactic Plane Survey 2nd Epoch (MGPS-2) Compact Source Catalogue \citep{b54}; a cut-off image of the source shows an unresolved but elongated structure. QSO B0241+62 is reported as a relatively strong source in various radio surveys, has a  flat radio spectrum with  a $\alpha$$\sim$0.1 \footnote{$S_{\nu}$$\propto$$\nu^{-\alpha}$} \citep{b62} and so almost certainly qualifies as a radio loud AGN. IGR J13109-5552 has less information at radio frequencies but the few available data  provide a flat spectrum also in this case ($\alpha$$\sim$0.3).

\begin{small}
\begin{figure*}
\centering
\includegraphics[width=0.3\linewidth,angle=-90]{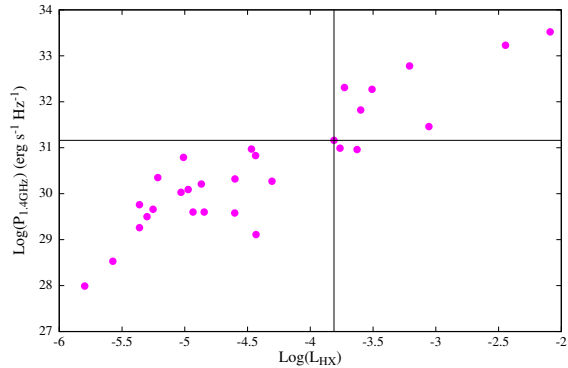}
\caption{1.4 GHz power density versus  the 1.4 GHz to 20-100 keV flux density ratio for  most sources in the \emph{INTEGRAL} BL AGN catalogue. The
boundaries on both axis are set by the values of the weakest FRII in the sample (IGR J21247+5058) and are defined by the lines drawn in the figure.} \label{radio_bright}
\end{figure*}
\end{small}

\begin{table*}
\tiny
\begin{center}
\centerline{{\bf Table 1}}
\vspace{0.2cm}
\begin{tabular}{lccc}
\hline
\multicolumn{4}{c}{{\bf Candidate Radio Loud AGN }}\\
\hline
{\bf Name}&{\bf Type} &{\bf Log(P$_{5GHz}$)}       & {\bf Log(R$_B$)}\\            &            &{\bf erg s$^{-1}$Hz$^{-1}$} &   \\
\hline
QSO B0241+62         & -     &  31.230 &  2.8-2.9\\
B3 0309+411          &  BLRG, FRII?   & 32.322 &   3.3 \\
3C 111               &  BLRG, FRII     &  31.919 &   4.4-5.0 \\
IGR J13109-5552      &  -        & 31.929 &   2.7 \\     3C 390.3             & BLRG, FRII    &  32.477 & 3.7 \\ 4C 74.26             &  BLRG, FRI/FRII&  31.939 &   1.9-2.0\\
S5 2116+81           &  BLRG, FR?     & 31.771 &   1.9 \\
IGR J21247+5058      &  BLRG, FRII   &  31.041 &   - \\
\hline
MCG+08-11-011         &   -     & 29.886 &  1.6 \\
Mrk 6                 &   -     & 29.934 &  1.5-1.8\\
\hline
\end{tabular} \end{center}
\scriptsize
\end{table*}

\begin{table*}
\tiny
\begin{center}
\centerline{{\bf Table 1A}}
\vspace{0.2cm}
\begin{tabular}{lccccccc}
\hline
\multicolumn{8}{c}{{\bf  Source Data}}\\
\hline
{\bf Name}      & {\bf RA}&{\bf Dec}& {\bf z} &{\bf N$_H^{gal}$}        & {\bf F$_{20-100keV}$}                 & {\bf L$_{20-100keV}$}       &{\bf CD$^{\ddagger}$}\\
                 &         &          &         &{\bf 10$^{22}$cm$^{-2}$}&{\bf 10$^{-11}$erg cm$^{-2}$s$^{-1}$}& {\bf 10$^{44}$erg s$^{-1}$}&   \\
\hline
QSO B0241+62    &  41.285 &  +62.480 &  0.044  &           0.75            & 6.34 & 2.63 & 1.94\\
B3 0309+411     &  48.273 &  +41.343 &  0.136  &           0.13            & 2.57 & 9.75 & 1.17 \\
3C 111          &  64.573 &  +38.014 &  0.0485 &           0.32            & 9.46 & 4.75 & 0.04\\
IGR J13109-5552v& 197.682 & -55.863  &  0.085  &           0.22            & 2.42 & 0.33 & - \\
3C 390.3        & 280.586 &  +79.781 &  0.0561 &           0.04            & 6.06 & 4.06 & 0.09\\ 4C 74.26        & 310.585 &  +75.145 &   0.104 &           0.12            & 4.87 & 10.97& 0.95\\
S5 2116+81      & 318.492 &  +82.072 &  0.084  &           0.07            & 4.06 & 6.02 & 1.78\\
IGR J21247+5058 & 321.172 &  +50.972 &  0.02   &           1.11            & 10.90& 0.94 & 0.76\\
\hline
\end{tabular}
\end{center}
\scriptsize
$^{\ddagger}$: radio core dominance at 5 GHz, except for B3 0309+411B for which it is measured at 8 GHz (see references in the text).\\
\end{table*}

We also checked, for completeness, the morphology of Markarian 6 and MCG+08-11-011 in the NVSS/NED database: both are similar to QSO B0241+62, i.e. display a compact core which, at higher resolution, reveals a clear (in Markarian 6; \citealt{b11}) or possible (in MCG+08-11-011; \citealt{b45}) double lobe structure.
However their radio spectra are steeper ($\alpha$$\sim$0.6-0.7; \citealt{b62}) than in the previous two cases and more similar to radio quiet AGN. This evidence, coupled with their lower radio power argues against a classification as radio loud AGN and therefore both objects are not considered further.
 
Table 1A lists all relevant information on the 8 sources chosen for this study, including redshift, Galactic column density in their direction, 20-100 keV flux and luminosity and the 5/8 GHz radio core dominance (CD). The Galactic column densities are taken from \citet{b5},
while the parameter \emph{CD=S$_{core}$/(S$_{tot}$-S$_{core}$)} is taken from \citet{b6}; the only exceptions are B3 0309+411 and IGR J21247+5058, for which the \emph{CD} parameter has been evaluated using data from \citet{b47} and \citet{b18} respectively. No information on CD is available for IGR J13109-5552.

The radio core dominance spans a large range of values from 0.04 to 1.9. Sources with high values of this parameter are those in which the beamed radio emission is emitted in a direction closer to the line of sight; in these sources, jet emission
is likely to play a greater role than in sources with smaller values of \emph{CD}. Contrary to expectations, we do not find any convincing
correlation between \emph{CD} and either (P$_{1.4GHz}$) or R$_{HX}$, indicating that in at least some of our sources a jet component alone cannot be the origin of the source radio loudness. As a final remark, we point out that our 8 radio loud sources make up around 20\% of the  20-40 keV complete sample of BL active galaxies and are therefore a non-negligible fraction of the \emph{INTEGRAL} AGN population.

\section{Data reduction}

The sources in our sample have all been observed by \emph{XMM-Newton} with the
exceptions of QSO B0241+62, for which \emph{Chandra} data are available, and IGR J13109-5552 and S5 2116+81, for which \emph{Swift/XRT} data have instead been used.  
MOS and pn \citep{b25,b26} data were reprocessed using the \emph{XMM-Newton} Standard
Analysis Software (SAS) version 7.0 employing the latest available calibration files. Only patterns corresponding to single, double, triple and quadruple
X-ray events for the two MOS cameras were selected (PATTERN$\leq$12), while
for the pn only single and double events (PATTERN$\leq$4) were taken
into account; the standard selection filter FLAG=0 was applied.  Observations have been filtered for periods of high background and the
resulting exposures are listed in table 2. Source counts were extracted from circular regions of typically 40$^{\prime\prime}$-50$^{\prime\prime}$ of radius centered on the source, while background spectra were extracted from circular regions close to the source or
from source-free regions of typically 20$^{\prime\prime}$ radius. The ancillary response matrices (ARFs) and the detector response matrices (RMFs) were
generated using the \emph{XMM}-SAS tasks \emph{arfgen} and \emph{rmfgen};
spectral channels were rebinned in order to achieve a minimum of 20 counts per each bin.
For sources affected by pile-up (3C 111, 4C 74.26 and 3C 390.3) the central 5$^{\prime\prime}$ of the
PSF have been excised and spectra have been extracted from annular regions of typically 50$^{\prime\prime}$ external radius. 
\emph{XRT} data reduction for IGR J13109-5552 and S5 2116+81 was performed using the XRTDAS v1.8.0
standard data pipeline package (\texttt{xrtpipeline} v. 0.10.3) in order
to produce screened event files. All data were collected in the
Photon Counting (PC) mode \citep{b9}, adopting the standard grade
filtering (0-12 for PC) according to the \emph{XRT} nomenclature.  Source data have been extracted using photons in a circular region of
radius 20$^{\prime\prime}$; background data have been taken
from various uncontaminated regions near the X-ray source,
using either a circular region of different radius or an annulus
surrounding the source.

\emph{Chandra} data reduction for QSO B0241+62 was performed with CIAO 3.4 and CALDB 3.2.4 to
apply the latest gain corrections. Subsequent filtering on event grade
and exclusion of periods with high background resulted in a total exposure of
34 ks. The CIAO script {\textquotedblleft{psextract}\textquotedblright} was used to generate the source spectrum, with its appropriate background and response files; spectral data were extracted from a circular region of radius  $\sim$ 4$^{\prime\prime}$ while background files were generated using a region of $\sim$18$^{\prime\prime}$ in diameter. With a count rate of 0.14 counts/frame, the pile-up fraction is insignificant ($<$10\%).

The \emph{INTEGRAL} data reported here consist of several pointings  performed by the \emph{IBIS/ISGRI} \citep{b28,b13}
instrument between revolutions 12 and 429, i.e. the period from launch to the end of April
2006. \emph{ISGRI} images for each available pointing were generated in various energy
bands using the ISDC offline scientific analysis software OSA \citep{b8} version 5.1. Count rates at the position of the source were extracted from individual images in order to
provide light curves in various energy bands; from these light curves, average fluxes were
then extracted and combined to produce an average source spectrum (see \citealt{b3} for details). Analysis
was performed in the 20-110 keV band. In the present analysis we also made use of public \emph{Swift/BAT} spectra, retrieved on the
web\footnote{http://swift.gsfc.nasa.gov/docs/swift/results/bs9mon/};
spectra are from the first 9 months of operations of the \emph{Swift/BAT} telescope \citep{b55}.

Table 2 reports the observation log of each source, i.e. the observation date of each X-ray observation and the exposure time for each instrument employed in the spectral analysis, with the exception of \emph{Swift/BAT}.

\begin{table*}
\scriptsize
\begin{center}
\centerline{{\bf Table 2}}
\vspace{0.2cm}
\begin{tabular}{lcccccc}
\hline
\multicolumn{7}{c}{{\bf Observations Log}}\\
\hline
{\bf Name}       & {\bf Obs. Date} & {\bf Exposure (pn)} & {\bf Exposure (MOS)} & {\bf Exposure (\emph{XRT})}&{\bf Exposure (\emph{Chandra})}& {\bf Exposure (\emph{INTEGRAL$^{\dagger}$})}\\
                        &                        &      {\bf (sec)}           &    {\bf (sec)}                &   {\bf (sec)}                           &           {\bf (sec)}                       &                     {\bf (sec)}                  \\
\hline
QSO B0241+62      &   06-07/04/2001 &   -     &        -         & - & 34000 &  371986 \\
B3 0309+411       &   04/09/2005    & 10990.1 &  19703.1/16138.6 & - &   -   &  506677 \\  3C 111            &   14/03/2001    & 14542.1 &  76262.3/7116.3  & - &   -   &  200566\\    IGR J13109-5552   &   25/12/2006    &    -    &      -           & 6.58 & - & 1288 \\
3C 390.3          &   08/10/2004    & 38762.3 &  32450.7/34603.9 & - &   -   &  305807 \\ 4C 74.26          &   06/02/2004    & 25643.5 &  28960.4/28960.4 & - &   -   &  121316 \\
S5 2116+81        &   25/06/2006    &    -    &       -          & 5111.4 &   -   &  198462 \\
                  &   17/10/2006    &    -    &       -          & 5237.4 &   -   &     -   \\
IGR J21247+5058   &   06/11/2005    & 22166   &  24245/23614     & - &   -   & 1067000 \\
\hline
\end{tabular}
\end{center}
\scriptsize
$^{\dagger}$: note that for \emph{INTEGRAL} the exposure refer to a number of pointings in the period between launch and April 2006.
\end{table*}

\section{Broad-band spectral analysis}

The \emph{XMM-Newton}, \emph{Chandra} and \emph{Swift/XRT} data were fitted together with \emph{INTEGRAL/ISGRI} and \emph{Swift/BAT} data using \texttt{XSPEC} v.11.2.3 \citep{b1}; errors are quoted at 90\% confidence level for one parameter of interest ($\Delta\chi^2$=2.71). In the fitting procedure, a multiplicative constant, \emph{C}, has been introduced to take into account possible cross-calibration mismatches between the X-ray and the soft gamma-ray data. When treating the \emph{INTEGRAL} data, this constant \emph{C$_1$} has been found to be close to 1 with respect to \emph{XMM-Newton}, \emph{Swift/XRT} and \emph{Chandra} using various source typologies \citep{b12,b4,b16,b21}, so that significant deviation from this value can be confidently ascribed to source flux variability; also when considering \emph{BAT}
data, a cross-calibration constant \emph{C$_2$} different to 1 often means flux variation in the source analysed (see for example \citealt{b56} and \citealt{b57}). We also introduced cross calibration constants between the XMM instruments, pn/MOS1 and MOS2/MOS1; these were left free to vary and always found to be in the range 0.97-1.04. In all our fits, Galactic absorption (see table 1A) has already been taken into account so that any column densities reported in subsequent tables refer to absorption intrinsic to the source.

\subsection{Towards a first approximation of the broad-band continuum.}

We firstly fitted the broad-band data in the 0.5-110 keV energy range employing a simple power law (absorbed only by Galactic column
density) in order to identify typical features of the AGN spectra.  Some of our objects show evidence of excess counts below 1
keV; however, since the study of the soft excess is not the main objective of the present work, but rather the understanding of the high energy emission characteristics
(photon index, high energy cut-off and reflection), we have focussed the analysis in the 1-110 keV energy range. Starting from the simple power law we introduced various
spectral features such as intrinsic absorption and iron line, each time performing an F-test to verify the statistical significance of each new component and to provide a basic model able to describe the data at least to a first approximation (see table 3 and residuals with respect to this model in figures \ref{pl}, \ref{pl1}, \ref{pl2} and
\ref{pl3}). 

Inspection of figure \ref{pl2} (left panel) suggests that the model employed in table 3 does not provide a good description of the data of 3C 390.3, where the residuals have
an unusual concave shape, which may be due to the presence of a low energy component (soft excess) still affecting the spectrum around 2-3 keV. To take this into account we
have added an extra feature to the model in the form of another power law or a blackbody component: both fits provide an equally significant improvement (typically more than
99\%) and more acceptable residuals. The blackbody temperature is slightly higher than typically observed in Seyfert 1 galaxies \citep{b21}. It is interesting to note that
so far no soft excess has been reported for this source except for an old claim never confirmed afterwards \citep{b38}; however a \emph{Chandra} image of 3C 390.3 reveals
extended soft X-ray emission around the nucleus so the soft excess photons we detect could be related to this component (\citealt{b61} and figure 7 in
\citealt{b39}). We also point out that this source is known to have a variable column density, as found by Grandi et al. (1999); however in our analysis we were not able to detect
any absorption in excess to the Galactic one. Due to the evidence found, in the following we adopt the model \texttt{wa$_g$*wa*(bb+po+zga)} (table 3A and figure \ref{21247_3c390_bb} right panel) as our basic representation of 3C 390.3.

It is also evident from table 3 and figure \ref{pl3} (right panel) that IGR J21247+5058, has a very flat power law slope and consequently shows a mismatch between X-ray and soft gamma-ray data. It is however known from the literature that this source requires a model that takes into account complex absorption \citep{b18} in the form of two layers of material partially covering the central source. Hence, in the case of IGR J21247+5058, the simple intrinsic photoelectric absorption has been substituted by two partial covering components (\texttt{wa$_g$*pcfabs*pcfabs*(po+zga)} in \texttt{XSPEC}, see results in table 3B); when this model is applied, the power law index becomes steeper and the X-ray/gamma-ray match is more acceptable (see figure \ref{21247_3c390_bb} left panel).

Using the results reported in table 3, 3A and 3B we can already draw some conclusions, as the models employed at this stage provide, to a first approximation, a good description of the broad-band spectra of all our sources. In all but 2 AGN (B3 0309+411 and 3C 390.3), an intrinsic absorption component is strongly required by the data,
at more than 90\% confidence level for IGR J13109-5552 and S5 2116+81 and at more than 99\% confidence level for the remaining objects.
Note, however, that in 3C 111 the measured absorption could be due to a molecular cloud located between us and the source \citep{b31}
and not intrinsic to the AGN. It is also interesting to note that only one out of eight objects displays complex absorption; a similar  fraction was found in a sample of 9 radio quiet Seyfert 1 galaxies discussed by \citet{b21}, where at least two objects required
one or two layers of cold material partially obscuring the central nucleus.

The Fe K$\alpha$ line is detected in all the sources (99\% confidence level), except in IGR J13109-5552 and S5 2116+81, but this may be due to \emph{XRT} not being sensitive enough to detect the cold iron line around 6.4 keV. The line is narrow in all of
our sources (hence the line width has been fixed to 10 eV), except in 4C 74.26, in which a quite strong and possibly broad iron line is detected (see also
\citealt{b2} and \citealt{b58}). The line equivalent widths (EW) are found to be $\lesssim$100 eV in all  of our AGN. In two sources, namely 3C 111 and IGR J21247+5058,
the equivalent width values are close to the capability limits of moderate resolution CCD instruments like \emph{XMM-Newton}, calling for some caution in distinguishing between a real feature or just local noise.
If the line is related to reflection, the observed values would imply a low reflection component parameter (R=$\Omega$/2$\pi$): in this case one may expect to find the EW to \emph{R} ratio of the order of 100-130 eV \citep{b22} and consequently values of \emph{R}$\leq$1. We will come back to this point later.

Finally a note on the cross-calibration constants between the X-ray data and \emph{IBIS/BAT} points: in some sources these constants are not consistent with 1, indicating, as expected, flux variations between the X-ray snap-shot observations and the time-averaged high energy spectral data.

\begin{small}
\begin{figure*}
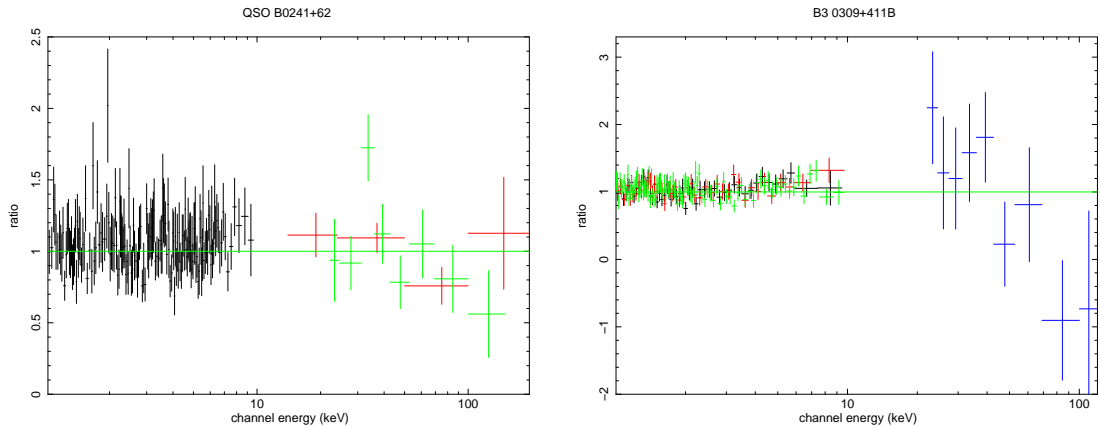

\centering
\includegraphics[scale=0.3,angle=-90]{qso0241_wawapozga_ra.ps}
\hspace{0.5cm}
\includegraphics[scale=0.3,angle=-90]{b0309_1_100_ra.ps}\\
\caption{Model to data ratios for QSO B0241+62 (left panel) and B3 0309+411B (right panel). The model employed is a simple power law absorbed
by both Galactic and intrinsic column density (only in the case of QSO B0241+62) plus a narrow Gaussian line component (see Table 3).}
\label{pl}
\end{figure*}
\end{small}

\begin{small}
\begin{figure*}
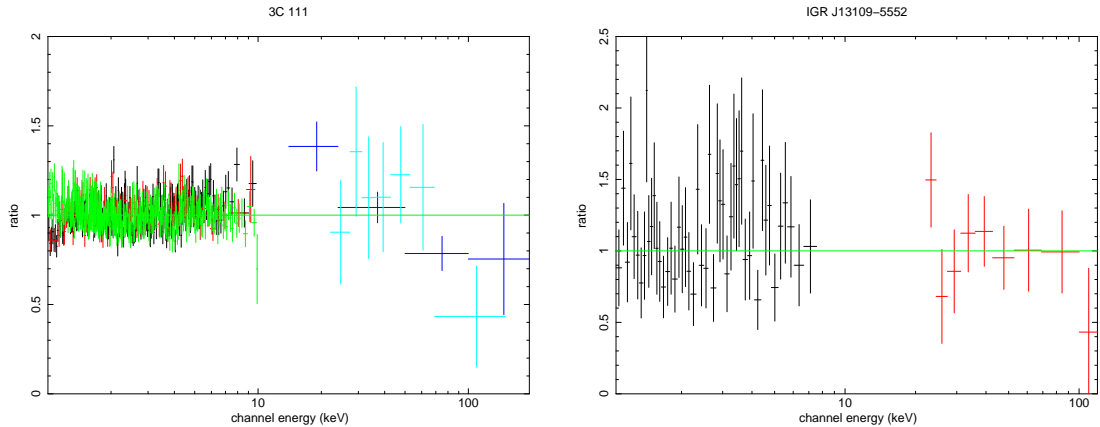

\centering
\includegraphics[scale=0.3,angle=-90]{3c111_wawapozga_ra.ps}
\hspace{0.5cm}
\includegraphics[scale=0.3,angle=-90]{igrj13109_wapo_ra.ps}\\
\caption{Model to data ratios for 3C 111 (left panel) and IGR J13109-5552 (right panel). The model employed is a simple power law absorbed
by both Galactic and intrinsic column density plus, in the case of 3C111, a narrow Gaussian line component (see Table 3).}
\label{pl1}
\end{figure*}
\end{small}

\begin{small}
\begin{figure*}
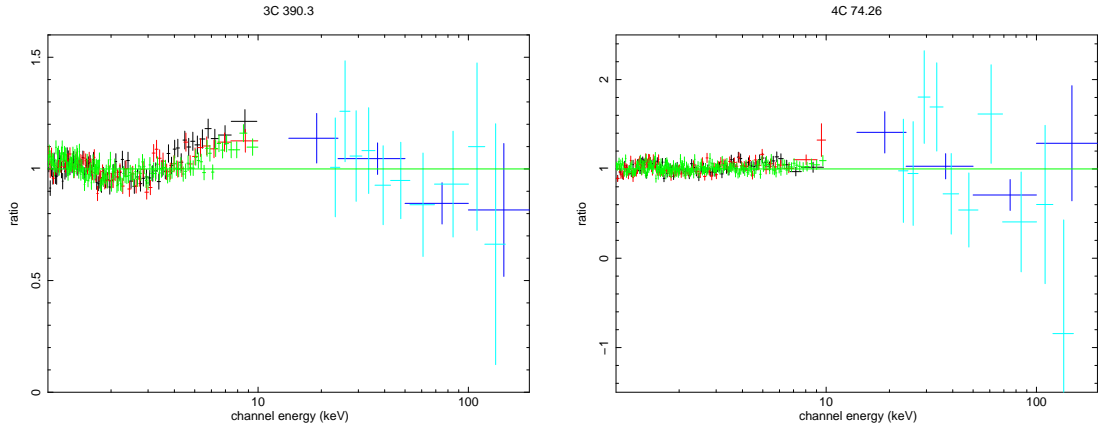

\centering
\includegraphics[scale=0.3,angle=-90]{3c390_bat_ra.ps}
\hspace{0.5cm}
\includegraphics[scale=0.3,angle=-90]{4c74_wawapozga.ps}\\
\caption{Model to data ratios for 3C 390.3 (left panel) and 4C 74.26 (right panel). The model employed is a simple power law absorbed
by both Galactic and intrinsic column density (only in the case of 4C 74.26) plus a narrow (3C 390.3) or broad (4C 74.26) Gaussian line component (see Table 3).}
\label{pl2}
\end{figure*}
\end{small}

\begin{small}
\begin{figure*}
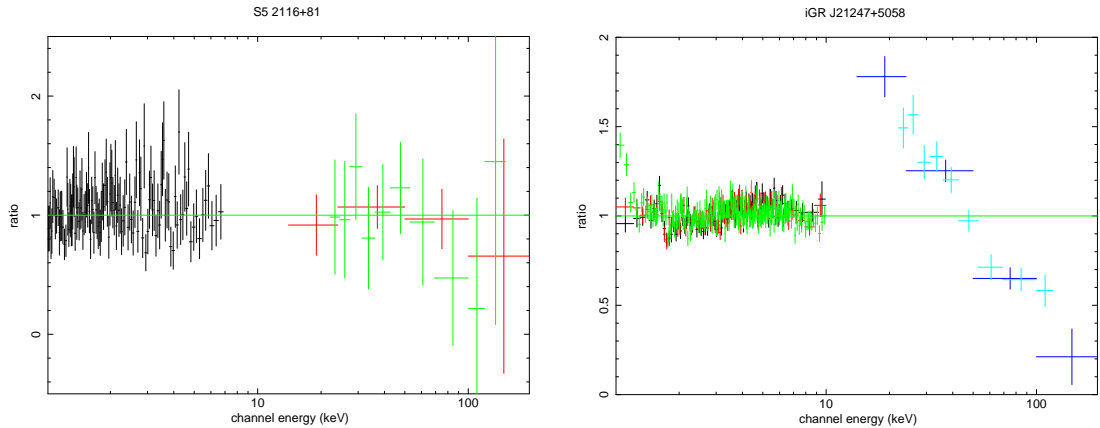

\centering
\includegraphics[scale=0.3,angle=-90]{s2116_bat_ra.ps}
\hspace{0.5cm}
\includegraphics[scale=0.3,angle=-90]{21247_wawapozga_ra.ps}\\
\caption{Model to data ratios for S5 2116+81 (left panel) and IGR J21247+5058 (right panel). The model employed is a simple power law absorbed
by both Galactic and intrinsic column density plus, in the case of IGR J21247+5058, a narrow Gaussian line component (see Table 3).}
\label{pl3}
\end{figure*}
\end{small}

\begin{small}
\begin{figure*}
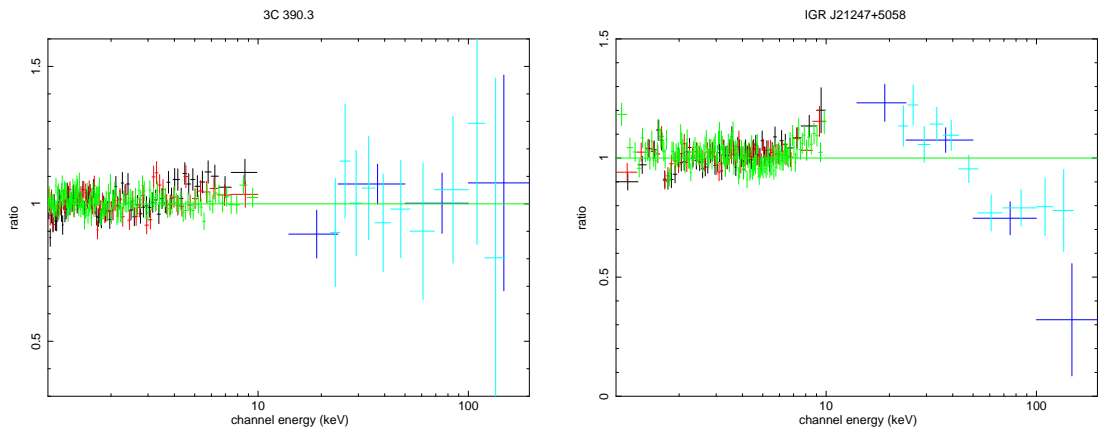

\centering
\includegraphics[scale=0.3,angle=-90]{3c390_bat_bb_ra.ps}
\hspace{0.5cm}
\includegraphics[scale=0.3,angle=-90]{21247_bat_pc_ra.ps}\\
\caption{Model to data ratio for 3C 390.3 (left panel) and IGR J21247+5058 (right panel).  The model used for 3C 390.3 is a simple power law absorbed by
Galactic column density plus a blackbody component to model the soft excess and a narrow Gaussian line component (Table 3A). For IGR J21247+5058 the model used is instead a simple power law absorbed by
Galactic column density and two layers of cold absorbing material partially covering the central source plus a narrow Gaussian line
component (Table 3B).}
\label{21247_3c390_bb}
\end{figure*}
\end{small}

\subsection{Incidence of the high energy cut-off.}

We next introduced a high energy cut-off in the primary power law of our basic model (see tables 4, 4A and 4B for model description in
each source case). Only two of the eight sources in the sample strongly ( at more than 99\% confidence level) require a cut-off energy, i.e. 3C 111 and IGR J21247+5058; the cut-off is localised at around 100 keV for 3C 111 and 80 keV for IGR J21247+5058.
As far as the other sources are concerned, we could only set a lower limit for their cut-off energies at around 60-80 keV,
apart from 3C 390.3, where the high energy cut-off could not be constrained at all. It is also interesting to note that the values of \emph{C$_{1}$} and \emph{C$_{2}$} tend to be higher when a cut-off energy component is added to the simple power law.

\subsection{Incidence of the reflection component.}

We then introduced reflection in our basic model of tables 3, 3A and 3B to verify the incidence of this spectral component in our
broad-band data (again refer to tables 5, 5A and 5B for model description in each source case).
We substituted the power law with the \texttt{pexrav} model in \texttt{XSPEC}, fixing the inclination angle at 30$^{\circ}$ (i.e. a nearly
face-on geometry as expected in type 1 AGN) and the cut-off energy at 10000 keV. Taking into consideration the fact that the iron line
EW is always $\lesssim$100 eV, we assume that the reflection parameter R cannot have very high values; for this reason we initially allowed the reflection component to vary in the range 0-2. B3 0309+411 and 4C 74.26 require reflection  at more than 99\% confidence level, while 3C 390.3 and QSO B0241+62 at a lower confidence level (98\% and  92\% respectively). The reflection component is also significantly required (at more than 99\% confidence level) in 3C 111 and IGR J21247+5058,
the two AGN which also have a high energy cut-off; since in these two objects the iron line is very weak or possibly even absent we expect to observe negligible or low reflection and we thus suspect that a better fit is achieved when both R and E$_{cut}$ are left free to vary (see next section).
In S5 2116+81 and IGR J13109-5552 the data only provide a loose lower limit on R, likely due to the low statistical quality of the X-ray data.

The reflection values are all above 0.1 and, when constrained, are typically around 1. Despite being highly required, R is unconstrained in B3 0309+411 due to the upper bound of 2 set on the reflection; when this bound is removed, R is found to be 2.44$^{+1.41}_{-1.17}$ and the constant becomes C$_{2}$=2.30$^{+1.10}_{-0.94}$. It is also
important to note that R tends to be higher than predicted from the measured iron line EW and that the \texttt{pexrav} model provides, as expected, smaller values of \emph{C$_{1}$} and \emph{C$_{2}$}  with respect to the previous two models.

\subsection{Constraining both reflection and cut-off.}

Finally, the broad-band spectra have been fitted with a cut-off power law reflected from neutral material (\texttt{pexrav} model but with E$_{cut}$ free to vary, see table 6 and figures \ref{qso_b3}, \ref{3c111_3c390} and \ref{4c_21247} for model description in each source case). We removed IGR J13109-5552 and S5 2116+81 from the sample, as in both sources the quality of the data allows no more complex fits than a simple power law (as in figure \ref{s5_13109}).
Because in B3 0309+411 the reflection parameter exceeds the value of 2, we raised the upper bound on R.
Although this model does not always provide a significant improvement with respect to previous fits, it nevertheless allows a simultaneous estimate of R and E$_{cut}$. We found that the reflection parameters of 5 objects and the cut-off energy of 4 are well
constrained; in the cases of B3 0309+411, 3C 111 and IGR J21247+5058, these represent the best fit results. Due to the above considerations, we employ in the following
discussion the values reported in table 6, with the eventual use of the upper limits on R and E$_{cut}$ obtained for IGR J13109-5552 and S5 2116+81. It is important to note
that the values of the parameters listed in table 6 are similar to those obtained in previous fits and also that in most sources the cross-calibration constants are now
compatible with 1, an indication that the model is appropriate and  that variability is not strong or common at high energies (see also \citealt{b36}). 
\section{Comparison with previous measurements}

Five out of eight sources in our sample have been previously studied over a broad energy range similar to ours, so that a direct comparison is possible. Overall, we find that our parameter values are in good agreement, or compatible within errors, with those found by these previous works. 
In the case of 3C 111, all observations  point to a small reflection and a compatibly low  iron line EW \citep{b7,b33}; our study yields a slightly higher value of R which is however  still
compatible, within errors, with previous results. We were able to put a stringent constraint on the high energy cut-off which we locate at a slightly lower value than in previous works, although well within the uncertainties found for example by \emph{BeppoSAX} \citep{b7}.

Comparison of our results for 3C 390.3 with the works by  \citet{b38} and  \citet{b34} indicates good agreement on the cut-off energy, although we, as Gliozzi et al., could not place a constraint on this parameter, which is instead found by Grandi and collaborators to be around 250 keV, i.e. still compatible with our lower limit of 300 keV. Our value of the reflection parameter is also fully compatible with the result obtained using BeppoSAX data \citep{b38}, but higher (although compatible within relative
uncertainties) than that obtained with RXTE spectra \citep{b34}.

We also find very good agreement with previous works for 4C 74.26 (e.g. \citealt{b32}, \citealt{b7} and \citealt{b58}), even though we find a slightly lower cut-off energy value. 
Not much can be said about S5 2116+81 because of the poor statistical quality of our data; however, comparison with \emph{BeppoSAX} data shows that our values do not contradict those found by \citet{b7}.

Finally our results are in full agreement with a previous analysis of X-ray/\emph{INTEGRAL} data of IGR J21247+5058 \citep{b18}. For QSO B0241+62, B3 0309+411 and IGR
J13109-5552 no previous work is available and thus the analysis presented here is the first  broad-band study for those 3 AGN.

\section{Discussion and Conclusions}

In the following we discuss the distribution of the fit parameters in our sample of AGN (see table 6) in comparison to a similar study made on a set of 9 radio quiet Seyfert
1 \citep{b21} also detected by \emph{INTEGRAL}. The power law slopes are found to be similar in both samples, i.e. around 1.5-2.0; only one of the radio bright AGN (the one
requiring complex absorption) shows a flat power law continuum, while low values of $\Gamma$ are more numerous in the sample of  \citet{b21}. The intrinsic absorption
measured in our sources is generally small or absent ($\lesssim$2$\times$10$^{21}$ atoms cm$^{-2}$), except for 3C 111 and IGR J21247+5058, and also similar to the values
found for radio quiet  type 1 AGN; in 3C 111 the extra absorption is probably due to intervening material between us and the source, while in the case of IGR J21247+5058
complex absorption is strongly required by the data. It is interesting to note that similar complexity, i.e. one or more layers of cold material partially covering the
central nucleus, is also observed in 2 of the 9 objects studied by \citet{b21}. We also point out that, up to very recently, the only other BLRG known to have a similar
spectral complexity regarding the absorption was 3C 445 \citep{b40}, for which three layers of cold material were required to model the spectrum.
Our analysis suggests that  objects with complex absorption are present within the population of broad line AGN, independent of them being radio quiet or radio loud. \citet{b7} found evidence in their data for a correlation between absorption and radio core dominance, with more absorbed objects showing lower
\emph{CD} values. We do not find such a trend; furthermore we do not have any indication of a correlation between photon index or absorption and R$_{HX}$ using both our
sources and the radio quiet Seyferts of \citet{b21}.

From the spectral analysis presented here, it is also evident that the high energy cut-off (although not required by all the sources 
in the sample) spans a wide range of values from $\sim$40 keV up to more than 300 keV, as already found for radio quiet AGN (see for instance \citealt{b35} and
\citealt{b21}). We also explored the possibility of a correlation between the high energy cut-off and the photon index but found nothing; a trend of increasing cut-off
energy with higher $\Gamma$ has been reported in the literature (see for instance \citealt{b22}), but due to the strong dependency between these two parameters in the
fitting procedure, it is difficult to discriminate between any true correlation and induced effects.

As far as the reflection fraction is concerned, we find in all but two cases (namely B3 0309+411 and 4C 74.26) very low values for this component, typically below 1; as
expected in these sources, the EW of the iron line  is $\lesssim$100 eV.
Indeed if the iron line emission is associated with the optically thick material of the disk, one would expect the line EW to correlate
with the reflection fraction as observed. However, in B3 0309+411 and 4C 74.26 the observed iron line EW is too small for the reflection measured: a possible explanation
resides in the scatter expected in the EW values due to a variation in the iron abundance, or to a possible anisotropy of the source seed photons which might affect the
observed spectrum \citep{b42,b43}. For a given value of \emph{R}, the EW could also differ according to the value of the power law photon index: up to $\Gamma$=2 the EW
decreases as the spectrum steepens, while above $\Gamma$=2 the trend is reversed \citep{b22, b41}. B3 0309+411 and 4C 74.26 both have steep spectra and so a lower EW than
expected on the basis of the measured \emph{R} value is not a surprise.

Overall we can conclude that, as already observed by a number of authors (i.e. \citealt{b7} and \citealt{b24}), the reprocessing features of ttthe radio loud AGN analysed here tend to be, on
average, quite weak. For a better comparison with their radio quiet counterparts, we have again used our data in combination with those of \citet{b21}. Figure \ref{r_ew}
shows a plot of \emph{R} vs. EW  for the entire sample. It is evident that radio loud AGN are more confined to a region of the plot characterized by low values of  EW and to a lesser extent of \emph{R}, whereas radio quiet objects tend to be more spread over both axes. This evidence does not seem to be related to the radio core dominance, suggesting that any dilution of the reprocessing features, at least for the sources reported here, is not caused by the presence of a jet.
Since the observed difference is not striking, an accretion flow origin for the X/gamma-ray emission is a likely explanation for the production of the reprocessing features in our radio loud AGN; this agrees well with some of them being classified as FR II sources, i.e. the BLRG most closely resembling radio quiet AGN \citep{b44}. Within this scenario, it is still possible that a different geometry and/or accretion flow efficiency involving the disk provides the condition for weaker reflection components in BLRG. Alternatively weaker  reflection features might be the result of reprocessing in an ionised accretion disk, as suggested by \citet{b60}; this would alleviate the need for a change in the accretion disk geometry and provide a more similar enviroment for both radio loud and radio quiet AGN.

Our sample is of course very small, limited to sources which have a small dynamic range of parameters (redshift, loudness, core dominance, etc.), and contains two objects
(i.e. QSO B0241+62 and IGR J13109-5552) which deserve further radio studies and are not immediately confirmed as BLRG.
We hope in the near future to improve the statistics, adding newly discovered AGN selected in the hard X-ray band in order to verify our findings as well as our novel
approach of selecting radio loud type 1 sources.

\begin{table*}
\scriptsize
\begin{center}
\centerline{{\bf Table 3}}
\vspace{0.2cm}
\begin{tabular}{lcccccccc}
\multicolumn{9}{c}{{\bf Spectral Fit Results: \texttt{wa$_g$*wa*(po+zga)}}}\\
\hline
{\bf Name}    &  {\bf N$_H$}               & {\bf $\Gamma$}&{\bf E$_{line}$}  & {\bf $\sigma$} &   {\bf EW}   & {\bf C$_1^A$} &{\bf C$_2^B$}&{\bf $\chi^2$ (dof)}\\
               &{\bf (10$^{22}$ cm$^{-2}$)}&                &    {\bf (keV)}   & {\bf (eV)}     & {\bf (eV)}   &                &              & \\
\hline
QSO B0241+62   &0.23$^{+0.08}_{-0.08}$&1.65$^{+0.07}_{-0.07}$&6.40$^{+0.06}_{-0.05}$&    10f        &85$^{+43}_{-45}$ &1.02$^{+0.22}_{-0.18}$&1.32$^{+0.32}_{-0.26}$& 227.1 (234) \\
B3 0309+411    &        -           &1.82$^{+0.03}_{-0.03}$&6.34$^{+0.07}_{-0.09}$&    10f        &100$^{+43}_{-50}$&      -             &4.65$^{+2.11}_{-2.00}$& 574.6 (603)\\
3C 111         &0.43$^{+0.02}_{-0.02}$&1.69$^{+0.02}_{-0.02}$&6.40$^{+0.05}_{-0.05}$&    10f        &  $<$30         &0.44$^{+0.05}_{-0.04}$&0.54$^{+0.11}_{-0.10}$& 1245.7 (1498) \\
IGR J13109-5552& 0.27$^{+0.23}_{-0.27}$ &1.70$^{+0.24}_{-0.22}$&         -          &     -         &       -        &     -              &2.39$^{+1.84}_{-0.99}$& 41.2 (55)\\
3C 390.3       &         -          &1.74$^{+0.01}_{-0.01}$&6.44$^{+0.02}_{-0.04}$&    10f        &  71$^{+8}_{-18}$&0.90$^{+0.07}_{-0.07}$&0.94$^{+0.10}_{-0.11}$& 2200.4 (2234)\\
4C 74.26       &0.12$^{+0.01}_{-0.01}$&1.73$^{+0.02}_{-0.01}$&6.44$^{+0.05}_{-0.06}$&183$^{+84}_{-65}$&103$^{+32}_{-25}$&0.54$^{+0.09}_{-0.09}$&0.87$^{+0.25}_{-0.25}$& 1992.0 (2060)\\
S5 2116+81     &    $<$0.19         &1.97$^{+0.12}_{-0.12}$&     -             &      -         &        -       &1.52$^{+0.60}_{-0.44}$&2.76$^{+1.34}_{-0.95}$& 89.1 (128) \\
IGR J21247+5058&0.69$^{+0.02}_{-0.02}$&1.33$^{+0.02}_{-0.01}$&6.39$^{+0.05}_{-0.08}$&    10f        &   $<$30        &0.32$^{+0.02}_{-0.02}$&0.36$^{+0.02}_{-0.02}$& 2692.3 (2559)\\
\hline
\end{tabular}
\end{center}
\scriptsize
\emph{A}: cross-calibration constant between \emph{XMM} and \emph{BAT}; \emph{B}: cross-calibration constant between \emph{XMM} and \emph{INTEGRAL/ISGRI}.
\end{table*}

\begin{table*}
\scriptsize
\begin{center}
\centerline{{\bf Table 3A}}
\vspace{0.2cm}
\begin{tabular}{lccccccc}
\multicolumn{8}{c}{{\bf 3C 390.3 Spectral Fits Results: \texttt{wa$_g$*(bb+po+zga)}}}\\
\hline
{\bf N$_H$}               &{\bf $\Gamma$} & {\bf kT}     &{\bf E$_{line}^{\dagger}$} & {\bf EW}  &{\bf C$_1^A$}& {\bf C$_2^B$}& {\bf $\chi^2$ (dof)}\\
{\bf (10$^{22}$cm$^{-2}$)}&                & {\bf (keV)} &    {\bf (keV)}            & {\bf (eV)}&              &               &  \\
\hline
           -            &1.89$^{+0.02}_{-0.02}$&2.41$^{+0.22}_{-0.17}$ &6.42$^{+0.04}_{-0.03}$& 49$^{+9}_{-15}$&1.37$^{+0.16}_{-0.15}$&1.58$^{+0.24}_{-0.21}$& 1970.4 (2232)\\
\hline
\end{tabular}
\end{center}
\scriptsize
$^{\dagger}$: line width fixed to 10 eV.\\
\emph{A}: cross-calibration constant between \emph{XMM} and \emph{BAT}; \emph{B}: cross-calibration constant between \emph{XMM} and \emph{INTEGRAL/ISGRI}.
\end{table*}

\begin{table*}
\scriptsize
\begin{center}
\centerline{{\bf Table 3B}}
\scriptsize
\vspace{0.2cm}
\begin{tabular}{lccccccccc}
\multicolumn{10}{c}{{\bf IGR J21247+5058 Spectral Fit Results: \texttt{wa$_g$*pcfabs*pcfabs*(po+zga)}}}\\
\hline
{\bf $\Gamma$}&{\bf N$_H^1$}                      &{\bf cf$_1$}& {\bf N$_H^2$}                     &{\bf cf$_2$}&{\bf E$_{line}^{\dagger}$}&{\bf EW}&{\bf C$_1^A$}&{\bf C$_2^B$}&{\bf $\chi^2$ (dof)}\\
                        &{\bf (10$^{22}$cm$^{-2}$)}&                   &{\bf (10$^{22}$ cm$^{-2}$)}&                    &    {\bf (keV)}    &{\bf (eV)} &          & & \\
\hline
1.73$^{+0.05}_{-0.04}$&10.81$^{+1.88}_{-1.49}$&0.38$^{+0.03}_{-0.03}$&1.08$^{+0.16}_{-0.15}$&0.86$^{+0.04}_{-0.03}$&6.39$^{+0.07}_{-0.08}$&$<$30&0.62$^{+0.06}_{-0.05}$&0.79$^{+0.08}_{-0.06}$& 2356.2 (2556)\\
\hline
\end{tabular}
\end{center}
\scriptsize
$^{\dagger}$: line width fixed to 10 eV.\\
\emph{A}: cross-calibration constant between \emph{XMM} and \emph{BAT}; \emph{B}: cross-calibration constant between \emph{XMM} and \emph{INTEGRAL/ISGRI}.
\end{table*}

\begin{table*}
\scriptsize
\begin{center}
\centerline{{\bf Table 4}}
\vspace{0.2cm}
\begin{tabular}{lcccccccccc}
\multicolumn{11}{c}{{\bf Spectral Fit Results: \texttt{wa$_g$*wa*(cutoffpl+zga)}}}\\
\hline
{\bf Name} &       {\bf N$_H$}        & {\bf $\Gamma$}& {\bf E$_{cut}$}      &{\bf E$_{line}$}         & {\bf $\sigma$}                 &   {\bf EW}       & {\bf C$_1^A$} &{\bf C$_1^B$}&{\bf $\chi^2$ (dof)}&{\bf Prob.$^{\dagger}$}\\
                            &{\bf (10$^{22}$ cm$^{-2}$)}&     &  {\bf (keV)}  &    {\bf (keV)}     & {\bf (eV)}                  & {\bf (eV)}       &  && &\\
\hline
QSO B0241+61    &0.18$^{+0.09}_{-0.09}$&1.58$^{+0.09}_{-0.09}$&$>$78&6.40$^{+0.07}_{-0.05}$&10f           &80$^{+41}_{-45}$&1.14$^{+0.23}_{-0.21}$&1.54$^{+0.39}_{-0.34}$& 222.9 (233)& 96\%\\
B3 0309+411     &-                   &1.80$^{+0.05}_{-0.05}$&   $>$43         &6.34$^{+0.07}_{-0.09}$&10f           &99$^{+46}_{-47}$&-&6.75$^{+3.44}_{-3.81}$&573.9 (602)& 91\%\\
3C 111          &0.42$^{+0.02}_{-0.02}$&1.65$^{+0.03}_{-0.03}$&110$^{+118}_{-40}$&6.40$^{+0.05}_{-0.05}$&10f           &$<$30           &0.56$^{+0.09}_{-0.08}$&0.74$^{+0.19}_{-0.17}$&1234.2 (1497)&99.9\% \\
IGR J13109-5552 &$<$0.46            &1.55$^{+0.31}_{-0.26}$&$>$58  &             -      &   -            &       -       &-&2.29$^{+1.84}_{-0.99}$& 39.8 (54)& 83\%\\  4C 74.26        &0.11$^{+0.01}_{-0.01}$&1.70$^{+0.03}_{-0.03}$& $>$78&6.44$^{+0.05}_{-0.06}$&184$^{+78}_{-66}$&104$^{+31}_{-26}$&0.63$^{+0.16}_{-0.14}$&1.09$^{+0.44}_{-0.37}$&1989.9 (2059)& 86\%\\
S5 2116+81      &$<$0.18 &1.95$^{+0.14}_{-0.14}$&$>$81&-&- &-&1.54$^{+0.64}_{-0.44}$&2.87$^{+1.59}_{-1.04}$&89.2 (127) & -\\
\hline
\end{tabular}
\end{center}
\scriptsize
\emph{A}: cross-calibration constant between \emph{XMM} and \emph{BAT}; \emph{B}: cross-calibration constant between \emph{XMM} and \emph{INTEGRAL/ISGRI}.\\
$^{\dagger}$: fit improvement with respect to table 3.
\end{table*}

\begin{table*}
\scriptsize
\begin{center}
\centerline{{\bf Table 4A}}
\vspace{0.2cm}
\begin{tabular}{lccccccccc}
\multicolumn{10}{c}{{\bf 3C390.3 Spectral Fits Results: \texttt{wa$_g$*(bb+cutoffpl+zga)}}}\\
\hline
{\bf N$_H$} & {\bf $\Gamma$}     & {\bf kT}             & {\bf E$_{cut}$}& {\bf E$_{line}^{\ddagger}$} & {\bf EW} & {\bf C$_1^A$}  & {\bf C$_2^B$}& {\bf $\chi^2$ (dof)}&{\bf Prob.$^{\dagger}$}\\
{\bf (10$^{22}$cm$^{-2}$)}&                     &{\bf (keV)}          & {\bf (keV)}    &  {\bf (keV)}               & {\bf (eV)}&           &       &     \\
\hline
    -          &1.89$^{+0.02}_{-0.01}$ & 2.41$^{+0.21}_{-0.16}$ &  NC        &6.43$^{+0.03}_{-0.04}$         & 48$^{+9}_{-14}$ & 1.37$^{+0.16}_{-0.15}$&1.58$^{+0.23}_{-0.11}$ & 1970.2 (2231)& 37\%\\
\hline
\end{tabular}
\end{center}
\scriptsize
$^{\ddagger}$: line width fixed to 10 eV.\\
\emph{A}: cross-calibration constant between \emph{XMM} and \emph{BAT}; \emph{B}: cross-calibration constant between \emph{XMM} and \emph{INTEGRAL/ISGRI}.\\
$^{\dagger}$: fit improvement with respect to table 3A.
\end{table*}

\begin{table*}
\scriptsize
\begin{center}
\centerline{{\bf Table 4B}}
\tiny
\vspace{0.2cm}
\begin{tabular}{lccccccccccc}
\multicolumn{12}{c}{{\bf IGR J21247+5058 Spectral Fit Results: \texttt{wa$_g$*pcfabs*pcfabs*(cutoffpl+zga)}}}\\
\hline
{\bf $\Gamma$}&{\bf N$_H^1$}                      &{\bf cf$_1$}& {\bf N$_H^2$}                     &{\bf cf$_2$}&{\bf E$_c$}&{\bf E$_{line}^{\ddagger}$}&{\bf EW}&{\bf C$_1^A$}&{\bf C$_2^B$}&{\bf $\chi^2$ (dof)}&{\bf Prob.$^{\dagger}$}\\
                       &{\bf (10$^{22}$cm$^{-2}$)}&                   &{\bf (10$^{22}$ cm$^{-2}$)}&                    & {\bf (keV)}&    {\bf (keV)}    &{\bf (eV)} &          & & &\\
\hline
1.49$^{+0.05}_{-0.07}$&0.80$^{+0.15}_{-0.17}$&0.88$^{+0.11}_{-0.05}$&8.23$^{+1.70}_{-2.05}$&0.27$^{+0.04}_{-0.05}$&80$^{+100}_{-63}$&6.39$^{+0.07}_{-0.08}$&$<$30&0.65$^{+0.05}_{-0.06}$&0.86$^{+0.07}_{-0.07}$& 2276.8 (2555)& $>$99.9\%\\
\hline
\end{tabular}
\end{center}
\scriptsize
$^{\ddagger}$ line width fixed to 10 eV.\\
\emph{A}: cross-calibration constant between \emph{XMM} and \emph{BAT}; \emph{B}: cross-calibration constant between \emph{XMM} and \emph{INTEGRAL/ISGRI}.\\
$^{\dagger}$: fit improvement with respect to table 3B.
\end{table*}

\begin{table*}
\scriptsize
\begin{center}
\centerline{{\bf Table 5}}
\vspace{0.2cm}
\begin{tabular}{lcccccccccc}
\multicolumn{11}{c}{{\bf  Spectral Fit Results: \texttt{wa$_g$*wa*(pexrav+zga)}, 0$\leq$R$\leq$2, E$_c$=10000}}\\
\hline
{\bf Name}      &  {\bf N$_H$}                        &  {\bf $\Gamma$}          &  {\bf R} & {\bf E$_{line}$} & {\bf $\sigma_{line}$} & {\bf EW}   & {\bf C$_1^A$} &{\bf C$_2^B$}&{\bf $\chi^2$ (dof)}&{\bf Prob.$^{\dagger}$}\\
                       &{\bf (10$^{22}$ cm$^{-2}$)}&                                     &              &    {\bf (keV)}     & {\bf (eV)}                  & {\bf (eV)}  &                        &                      & &\\
\hline
QSO B0241+61    & 0.31$^{+0.04}_{-0.10}$&1.83$^{+0.07}_{-0.15}$&$>$0.37&6.40$^{+0.08}_{-0.03}$&10f&67$^{+40}_{-42}$&0.59$^{+0.28}_{-0.11}$&0.74$^{+0.36}_{-0.16}$& 224.1 (233)& 92\%\\
B3 0309+411     &     -               &1.92$^{+0.03}_{-0.04}$&$>$1.20&6.33$^{+0.09}_{-0.10}$&10f & 70$^{+41}_{-43}$ &-&2.42$^{+1.06}_{-0.97}$&560.4 (602)& 99.99\%\\               3C 111          &0.48$^{+0.03}_{-0.03}$&1.78$^{+0.05}_{-0.06}$&1.06$^{+0.63}_{-0.68}$&6.40$^{+0.07}_{-0.08}$&10f&$<$30&0.30$^{+0.07}_{-0.04}$&0.35$^{+0.11}_{-0.07}$& 1238.5 (1497)&99.7\%\\
IGR J13109-5552 &0.26$^{+0.25}_{-0.25}$ &1.75$^{+0.23}_{-0.25}$&$>$0.1               &     -              &   -  &   -          &-&1.30$^{+2.69}_{-0.74}$& 40.5 (54)& 66\%\\
4C 74.26        &0.16$^{+0.02}_{-0.02}$&1.84$^{+0.05}_{-0.04}$&1.30$^{+0.67}_{-0.51}$&6.45$^{+0.04}_{-0.06}$&$<$186&51$^{+37}_{-16}$&0.33$^{+0.07}_{-0.06}$&0.54$^{+0.16}_{-0.17}$& 1977.0 (2059)&99.99\%\\
S5 2116+81      &$<$0.22 &2.03$^{+0.20}_{-0.16}$ &$>$0.1&-&- & -&1.10$^{+0.73}_{-0.43}$&1.97$^{+1.51}_{-0.87}$& 88.5 (127)&64\%\\
\hline
\end{tabular}
\end{center}
\scriptsize
\emph{A}: cross-calibration constant between \emph{XMM} and \emph{BAT}; \emph{B}: cross-calibration constant between \emph{XMM} and \emph{INTEGRAL/ISGRI}.\\
$^{\dagger}$: fit improvement with respect to table 3.
\end{table*}

\begin{table*}
\scriptsize
\begin{center}
\centerline{{\bf Table 5A}}
\vspace{0.2cm}
\begin{tabular}{lccccccccc}
\multicolumn{10}{c}{{\bf 3C 390.3 Spectral Fits Results: \texttt{wa$_g$*wa*(bb+pexrav+zga)}, 0$\leq$R$\leq$2, E$_c$=10000}}\\
\hline
{\bf N$_H$} & {\bf $\Gamma$}     & {\bf kT}             &  {\bf R}            &  {\bf E$_{line}^{\ddagger}$} & {\bf EW}            & {\bf C$_1^A$}  &{\bf C$_2^B$}&  {\bf $\chi^2$ (dof)}&{\bf Prob.$^{\dagger}$}\\
{\bf (10$^{22}$cm$^{-2}$)}&             & {\bf (keV)}         &                         &  {\bf (keV)}                         & {\bf (eV)}            &            &                &\\
\hline
  -     &1.89$^{+0.03}_{-0.02}$ & 2.38$^{+0.27}_{-0.22}$ & 0.70$^{+0.48}_{-0.59}$ & 6.43$^{+0.03}_{-0.04}$ & 44$^{+10}_{-13}$ & 0.88$^{+0.42}_{-0.13}$ & 0.93$^{+0.55}_{-0.22}$&1965.4 (2231)& 98\%\\
\hline
\end{tabular}
\end{center}
\scriptsize
$^{\ddagger}$: line width fixed to 10 eV.\\
\emph{A}: cross-calibration constant between \emph{XMM} and \emph{BAT}; \emph{B}: cross-calibration constant between \emph{XMM} and \emph{INTEGRAL/ISGRI}.\\
$^{\dagger}$: fit improvement with respect to table 3A.
\end{table*}

\begin{table*}
\scriptsize
\begin{center}
\centerline{{\bf Table 5B}}
\tiny
\vspace{0.2cm}
\begin{tabular}{lccccccccccc}
\multicolumn{12}{c}{{\bf IGR J21247+5058 Spectral Fit Results: \texttt{wa$_g$*pcfabs*pcfabs*(pexrav+zga)}, 0$\leq$R$\leq$2, E$_c$=10000}}\\
\hline
{\bf $\Gamma$}&{\bf N$_H^1$}                      &{\bf cf$_1$}& {\bf N$_H^2$}                     &{\bf cf$_2$}& {\bf R}& {\bf E$_{line}^{\ddagger}$}&{\bf EW}&{\bf C$_1^A$}& {\bf C$_2^B$}&{\bf $\chi^2$ (dof)}&{\bf Prob.$^{\dagger}$}\\
                       &{\bf (10$^{22}$cm$^{-2}$)}&                   &{\bf (10$^{22}$ cm$^{-2}$)}&                    &               & {\bf (keV)}    &{\bf (eV)} &        &   &\\
\hline
1.77$^{+0.05}_{-0.05}$&8.14$^{+1.81}_{-1.69}$&0.36$^{+0.03}_{-0.04}$&0.99$^{+0.16}_{-0.18}$&0.88$^{+0.05}_{-0.03}$&0.93$^{+0.59}_{-0.45}$ &6.38$^{+0.08}_{-0.08}$&$<$30&0.41$^{+0.08}_{-0.06}$&0.50$^{+0.10}_{-0.08}$& 2341.7 (2555)& 99.99\% \\
\hline
\end{tabular}
\end{center}
\scriptsize
$^{\ddagger}$: line width fixed to 10 eV.\\
\emph{A}: cross-calibration constant between \emph{XMM} and \emph{BAT}; \emph{B}: cross-calibration constant between \emph{XMM} and \emph{INTEGRAL/ISGRI}.\\
$^{\dagger}$: fit improvement with respect to table 3.
\end{table*}

\begin{table*}
\scriptsize
\centering
\begin{center}
\centerline{{\bf Table 6}}
\vspace{0.2cm}
\begin{tabular}{lcccccccc}
\multicolumn{9}{c}{{\bf  Spectral Fit Results: \texttt{wa$_g$*wa*(pexrav+zga)}}}\\
\hline
{\bf Name}     &{\bf N$_H$}                      &{\bf $\Gamma$}&{\bf R}&{\bf E$_c$}&{\bf EW$^{\dagger}$}  &{\bf C$_1^A$}&{\bf C$_2^B$}&{\bf $\chi^2$ (dof)}\\
                     & {\bf (10$^{22}$ cm$^{-2}$}&                       &            &{\bf (keV)}&{\bf (eV)}                     &                     &                       &           \\
\hline
QSO B0241+61    &0.21$^{+0.15}_{-0.10}$&1.64$^{+0.22}_{-0.14}$& 0.56$^{+0.82}_{-0.41}$ &    $>$86       &73$^{+44}_{-42}$& 0.85$^{+0.44}_{-0.32}$&1.12$^{+0.65}_{-0.48}$&222.2 (233)\\
B3 0309+411     &  -    & 1.90$^{+0.08}_{-0.08}$&3.48$^{+2.24}_{-1.58}$&35$^{+91}_{-17}$&59$^{+42}_{-43}$&-&6.89$^{+9.39}_{-3.90}$&554.2 (601)\\          3C 111          &0.46$^{+0.03}_{-0.03}$&1.73$^{+0.06}_{-0.06}$&0.85$^{+0.57}_{-0.58}$&126$^{+193}_{-50}$&    $<$30     &0.40$^{+0.11}_{-0.08}$&0.52$^{+0.19}_{-0.13}$&1227.9 (1497)\\
3C 390.3&    -                                       &1.89$^{+0.03}_{-0.02}$ &0.60$^{+0.60}_{-0.44}$ &$>$300&41$^{+12}_{-11}$&0.95$^{+0.17}_{-0.25}$&1.01$^{+0.39}_{-0.30}$& 1966.3 (2231)\\
4C 74.26        &0.14$^{+0.02}_{-0.03}$&1.79$^{+0.06}_{-0.07}$&1.22$^{+0.69}_{-0.70}$&100$^{+680}_{-52}$&88$^{+23}_{-20}$&0.95$^{+0.17}_{-0.25}$&1.01$^{+0.39}_{-0.30}$ & 1976.0 (2060)\\
IGR J21247+5058& complex$^{\star}$&1.48$^{+0.06}_{-0.06}$&$<$0.21&79$^{+23}_{-15}$&$<$30&0.65$^{+0.05}_{-0.08}$&0.86$^{+0.08}_{-0.12}$& 2276.7 (2555) \\
\hline
\end{tabular}
\end{center}
\scriptsize
$^{\dagger}$: line parameters fixed at value obtained in table 3.\\
\emph{A}: cross-calibration constant between \emph{XMM} and \emph{BAT}; \emph{B}: cross-calibration constant between \emph{XMM} and \emph{INTEGRAL/ISGRI}\\
$^{\star}$: N$_H^1$=7.86$^{+2.02}_{-1.66}$, cf$_1$=0.27$^{+0.04}_{-0.05}$; N$_H^2$=0.77$^{+0.18}_{-0.13}$, cf$_2$=0.89$^{+0.10}_{-0.06}$.
\end{table*}

\begin{small}
\begin{figure*}
\centering
\includegraphics[scale=0.3,angle=-90]{qso0241_pex_bat_uf.ps}
\hspace{0.5cm}
\includegraphics[scale=0.3,angle=-90]{b0309_pex_uf.ps}\\
\caption{Table 6 model for QSO B0241+62 (left panel) and B3 0309+411 (right panel). The model is a cut-off power law absorbed both by Galactic
and intrinsic column density  reflected by neutral material plus a narrow
Gaussian component describing the iron line.}
\label{qso_b3}
\end{figure*}
\end{small}

\begin{small}
\begin{figure*}
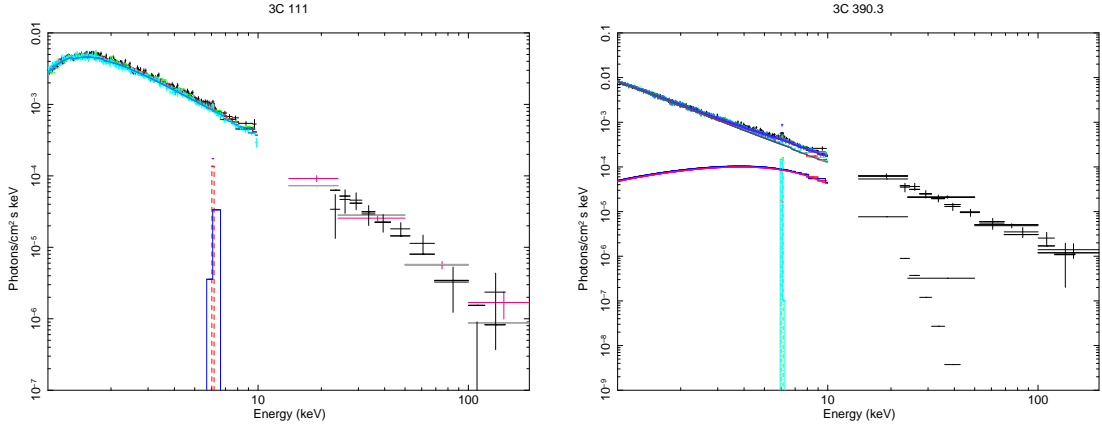

\centering
\includegraphics[scale=0.3,angle=-90]{3c111_pex_bat_uf.ps}
\hspace{0.5cm}
\includegraphics[scale=0.3,angle=-90]{3c390_pex_bat_uf.ps}\\
\caption{Table 6 model for 3C 111 (left panel) and 3C390.3 (right panel). The model is a cut-off power law absorbed both by Galactic and intrinsic column density reflected by neutral material plus a narrow Gaussian component describing the iron line.}
\label{3c111_3c390}
\end{figure*}
\end{small}

\begin{small}
\begin{figure*}
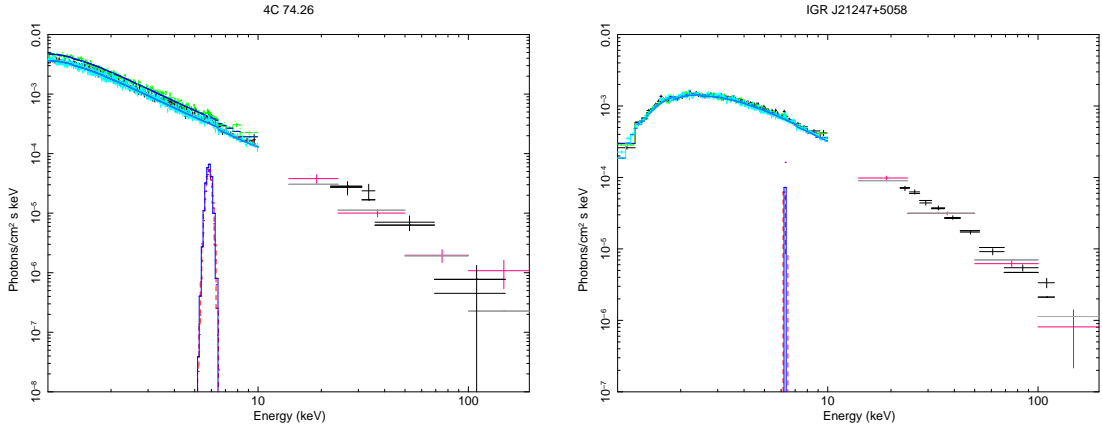

\centering
\includegraphics[scale=0.3,angle=-90]{4c74_pex_bat_uf.ps}
\hspace{0.5cm}
\includegraphics[scale=0.3,angle=-90]{21247_pex_bat_uf.ps}\\
\caption{Table 6 model for 4C 74.26 (right panel) and  IGR J21247+5058 (left panel). The model is a cut-off power law absorbed both by Galactic and intrinsic column density  reflected by neutral material plus a narrow Gaussian component describing the iron line.}
\label{4c_21247}
\end{figure*}
\end{small}

\begin{small}
\begin{figure*}
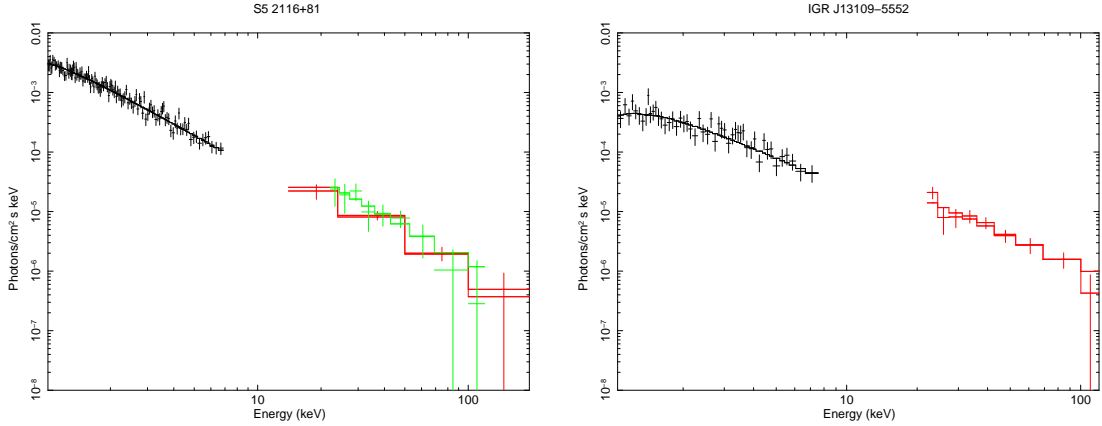

\centering
\includegraphics[scale=0.3,angle=-90]{s2116_po_bat_uf.ps}
\hspace{0.5cm}
\includegraphics[scale=0.3,angle=-90]{13109_po_uf.ps}\\
\caption{Table 3 model for S5 2116+81 (left panel) and IGR J13109-5552 (right panel). The model is a simple power law, absorbed both by Galactic and intrinsic column densities.}
\label{s5_13109}
\end{figure*}
\end{small}

\begin{small}
\begin{figure*}
\centering
\includegraphics[scale=0.4,angle=-90]{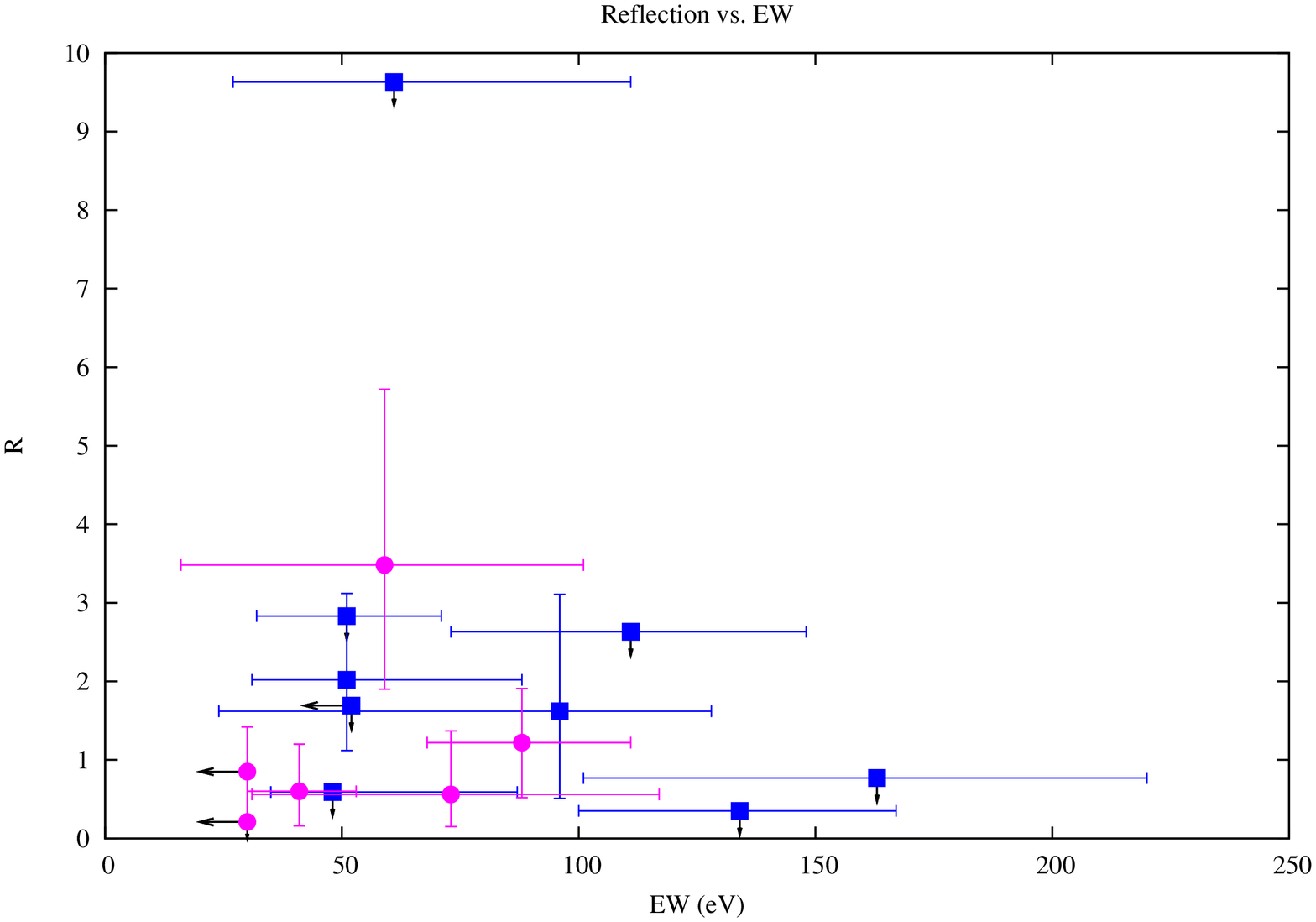}
\caption{Reflection fraction vs. EW for the sample analysed here and the sources presented in \citet{b21}. Blue squares are radio quiet sources, while radio loud sources are represented by magenta circles. Arrows represent upper or lower limits on the parameter values.}
\label{r_ew}
\end{figure*}
\end{small}

\section*{Acknowledgements}
We aknowledge the University of Southampton for financial support (M.M.) and the Italian Space Agency (ASI) financial
and programmatic support via contracts I/008/07/0 and I/088/06/0.
M. Molina wishes to thank V.A. McBride (Southampton University) for reducing \emph{Chandra} data for QSO B0241+62.


\label{lastpage}
\end{document}